\documentclass[12pt]{article}
\usepackage{graphicx}
\usepackage{url}

\title{Lagrangian tools to monitor transport and mixing in the ocean}
\author{S. V. Prants\thanks{E-mail: prants@poi.dvo.ru, \protect\url{http://dynalab.poi.dvo.ru}},
 M. V. Budyansky and M. Yu. Uleysky}
\date{\small Pacific Oceanological Institute of the Russian Academy of Sciences,\\ Vladivostok, 690041, Russia}

\begin{document}
\maketitle
\begin{abstract}
We apply the Lagrangian approach to study surface transport and mixing
in the ocean. New tools have been developed to track the motion of water masses,
their origin and fate and to quantify transport and mixing.
To illustrate the methods used we compute the Lagrangian
synoptic maps a comparatively small
marine bay, the Peter the Great Bay in the Japan Sea near Vladivostok city (Russia),
and in a comparatively large region in the North Pacific, the Kuroshio Extension
system. In the first case we use velocity data from a Japan Sea
circulation numerical model and in the second one the velocity data are
derived from satellite altimeter measurements of anomalies of the sea height
distributed by AVISO.
\end{abstract}


\section{Introduction}\label{aba:sec1}
The ocean presents a variety of dynamical phenomena with different space
scales ranging from millemeters to a few thousand of kilometers.
Despite of that, large-scale coherent
structures are easlily visible, say, at satellite images of the sea color and surface
temperature and can be identified
by means of in-situ measurements. The striking examples are the major western
boundary oceanic currents, the Gulf Stream in the Atlantic and the Kuroshio in
the Pacific. They are ``rivers'' with the warm water in the ocean with the width
on the order of 100--200~km and the maximal speed of current at the surface of 2 m/s.
Such currents separate waters with different physical, chemical and biological characteristics.
The other examples are mesoscale (with the size of a few hundred of kilometers)
and submesoscale (a few tenth of kilometers) eddies that can transport water over
hundreds and even thousands of kilometers and can survive for months
before breaking down. Being coherent features, they do not contain the same
waters but exchange them with the surrounding ocean, the process known as mixing.

Lagrangian and dynamical systems methods have been developed to study large-scale
transport and mixing in the ocean \cite{Haller,KP06,H02,OI09,LC05,MS06,WA06,PNAS}.
The main purposes of those studies are to track the fluid motion, to elucidate
and quantify transport and mixing processes. Simply speaking, we would like to
know where these or those waters come from, what is their fate and how they mix
in this or that region. In the Lagrangian approach one integrates trajectories
for a large number of synthetic particles advected by an Eulerian velocity field
\begin{equation}
\frac{d \vec r}{d t}=\vec v(\vec r, t).
\label{eq}
\end{equation}
The velocity field, $\vec v(\vec r, t)$, is supposed to be known
analytically, numerically or estimated from satellite altimetry.
While in the Eulerian approach we get frozen snapshots
of data, Lagrangian diagnostics enable to quantify spatio-time variability
of the velocity field. It has been established theoretically and experimentally
that even a simple deterministic velocity field may cause practically
unpredictable particle trajectories, the phenomenon known as chaotic
advection \cite{Ottino,KP06}. The real oceanic flows are not, of course,
deterministic and regular, but if the Eulerian correlation time is large
as compared to the Lagrangian one, the problem may be treated in the
framework of chaotic advection concept.

It is important to separate chaotic and turbulent mixing in the ocean.
The process of chaotic advection provides transport and mixing
with the characteristic scales on the order of a few tenths or even
hundreds of kilometers, whereas turbulence works at smaller scales.
At a comparatively large scale, turbulent mixing is homogeneous whereas
the chaotic one is not. Typical patterns of chaotic advection consist of
large-scale convoluted curves visible in some surface-temperature and
color satellite images. The effect of turbulent mixing is in small-scale
fluctuations superimposed on the large-scale convoluted curves.
If the velocity field on comparatively large scales is quasicoherent in space
and quasiregular in time but the motion of tracers is mainly irregular,
one deals with  chaotic mixing. Turbulent mixing means that the velocity
field is irregular in space and time at the same scales at which
the tracer's motion is irregular.

In this paper we report on our recent results on developing Lagrangian tools
to monitor surface transport and mixing in the ocean. We propose with this aim new
Lagrangian criteria that enable to track and quantify the water exchange processes
and reveal the underlying physical mechanisms. As an output, we compute different
Lagrangian synoptic maps of the regions under study for a given period of year
and analyze them. The methos is illustrated with a comparatively small
marine bay, the Peter the Great Bay in the Japan Sea near Vladivostok (Russia),
and a comparatively large region in the North Pacific, the Kuroshio Extension
system.  In the first case we have used velocity data from a Japan Sea
eddy-resolved circulation numerical model with the fine resolution of 2.5~km,
in the second one --- satellite altimetric velocity data with the coarse
resolution of the order of 35~km.

\section{Lagrangian and dynamical systems methods to study transport and mixing in the ocean}

Motion of a fluid particle in a two-dimensional flow is the trajectory
of a dynamical system with  given initial conditions governed by the velocity
field computed either by solving the corresponding master equations or as
the output of a numerical ocean model or derived from a measurement
\begin{equation}
\frac{d x}{d t}= u(x,y,t),\quad \frac{d y}{d t}= v(x,y,t),
\label{adveq}
\end{equation}
where $(x,y)$ is the location of the particle, $u$ and $v$ are the zonal
and meridional components of its velocity. Even if the Eulerian velocity
field is fully deterministic, the particle's trajectories may be very
complicated and practically unpredictable. It means that a distance
between two initially nearby particles grows exponentially in time
\begin{equation}
\| \delta {\mathbf r}(t) \| = \| \delta {\mathbf r}(0) \|\, e^{\lambda t},
\label{Lyap}
\end{equation}
where $\lambda$ is a positive number, known as the Lyapunov exponent,
which characterizes asymptotically the average rate
of the particle dispersion, and $\|\cdot\|$ is a norm of the vector
$\mathbf{r}=(x,y)$. It immediately follows from (\ref{Lyap}) that we are unable
to forecast the fate of the particles beyond the so-called predictability
horizon
\begin{equation}
T_p\simeq\frac{1}{\lambda}\ln\frac{\|\Delta \|}{\|\Delta (0)\|},
\label{horizon}
\end{equation}
where $\|\Delta  \|$ is the confidence interval of the particle location
and $\|\Delta (0)\|$ is a practically inevitable inaccuracy in
specifying the initial location. The deterministic dynamical system
(\ref{adveq}) with a positive maximal Lyapunov exponent for almost all
vectors $\delta \mathbf{r} (0)$ (in the sense of nonzero measure) is called
chaotic. It should be stressed that the dependence of the predictability
horizon $T_p$ on the lack of our knowledge of exact location is logarithmic,
i.~e., it is much weaker than on the measure of dynamical instability
quantified by $\lambda$. Simply speaking, with any reasonable degree of
accuracy on specifying initial conditions there is a time interval beyond
which the forecast is impossible, and that time may be rather short for
chaotic systems.

Since the phase plane of the two-dimensional dynamical system
(\ref{adveq}) is the physical space for fluid particles, many abstract
mathematical objects from dynamical systems theory (stationary points,
KAM tori, stable and unstable manifolds, periodic and chaotic orbits, etc.)
are material surfaces, curves and points in fluid flows. It is well known that
besides ``trivial'' elliptic fixed points, the motion around which is stable, there
are hyperbolic fixed points which organize fluid motion in their neighbourhood
in a specific way. In a steady flow the hyperbolic points are typically
connected by the separatrices which are their stable and unstable
invariant manifolds. In a time-periodic flow the hyperbolic points are replaced
by the corresponding hyperbolic trajectories with associated invariant manifolds
which in general intersect transversally resulting in a complex manifold
structure known as a heteroclinic tangle. The fluid motion in these regions
is so complicated that it may be strictly called chaotic,
the phenomenon known as chaotic advection \cite{Ottino,KP06}.
Adjacent  fluid particles in such tangles rapidly diverge providing
very effective mechanism for mixing.

Stable and unstable manifolds
are important organizing structures in the flow because they attract and repel
fluid particles (not belonging to them) at an exponential rate and
partition the flow into regions with different types of motion.
Invariant manifold in a two-dimensional flow is a material line, i.~e.,
it is composed of the same fluid particles in course of time.
By definition stable  ($W_s$) and unstable ($W_u$) manifolds of a hyperbolic
trajectory $\gamma(t)$~are material lines consisting of a set of points
through which at time moment $t$ pass trajectories asymptotical to
$\gamma(t)$ at $t \to \infty$ ($W_s$) and $t \to -\infty$ ($W_u$).
They are complicated  curves infinite in time and space that act
as boundaries to fluid transport.

The real oceanic flows are not, of course, strictly time-periodic.
However, in aperiodic flows there exist under some
mild conditions hyperbolic points and trajectories of a transient nature.
In aperiodic flows it is possible to identify aperiodically
moving hyperbolic points with stable and unstable effective manifolds
\cite{Haller,H02}. Unlike the manifolds in steady and periodic flows, defined
in the infinite time limit, the ``effective'' manifolds of aperiodic
hyperbolic trajectories have a finite lifetime. The point is that they
play the same role in organizing oceanic flows as do invariant manifolds
in simpler flows. The effective manifolds in course of their life undergo
stretching and folding at progressively small scales and intersect each other
in the homoclinic points in the vicinity of which fluid particles move
chaotically. Trajectories of initially close fluid particles diverge rapidly
in these regions, and particles from other regions appear there. It is the
mechanism for effective transport and mixing of water masses in the ocean.
Moreover, stable and unstable effective manifolds constitute Lagrangian
transport barriers between different regions because they are material
invariant curves that cannot be crossed by purely advective processes.

The stable and unstable manifolds of influencial hyperbolic trajectories
are so important because (1) they form a kind of a sceleton in oceanic flows,
(2) they divide a flow in dynamically different regions,
(3) they are in charge of forming an inhomogeneous mixing with
spirals, filaments and intrusions, (4) they are transport barriers
separating water masses with different characteristics.
Stable manifolds act as repellers for surrounding waters
but unstable ones are attractors. That is why unstable manifolds may be
rich in nutrients being oceanic ``dining rooms''.

There is a quantity, the finite-time Lyapunov exponents
(FTLE), that enables to detect and visualize stable and unstable manifolds
in complex velocity fields. The FTLE is the finite-time average
of the maximal separation rate for a pair of
neighbouring advected particles which is given by  \cite{OM}
\begin{equation}
\lambda (\mathbf{r}(t))\equiv\frac{1}{\tau}\ln\sigma (G(t)),
\label{Lyapunov}
\end{equation}
where $\tau$ is an integration time, $\sigma (G(t))$ the
largest singular value of the evolution matrix for linearized advection
equations. Scalar field of the FTLE is Eulerian but the very
quantity is a Lagrangian one that measures an integrated separation between trajectories.
Ridges (curves of the local maxima) of the FTLE field visualize
stable manifolds when integrating advection equations forward in time
and unstable ones when integrating them backward in time.

\section{Transport and mixing in marine bays}

When studing transport and mixing in marine bays, it is important to know
which waters enter the bay under study, which ones quit the bay, by which transport
corridors they do that and how the different waters mix in the bay interior.
The Lagrangian approach, allowing to compute the origin and fate of different
waters, is the most suitable for that.
Transport and mixing in marine bays is more inhomogeneous as compared with
those processes in open basins because of a complicated structure of currents
and eddies of different scales, strong tides and presence of river estuaries.
In this section we apply Largangian tools to
characrerize horizontal subsurface transport and mixing in the Peter the Great
Bay near Vladivostok city (Russia). That is the largest bay in the Japan Sea
with a few shallow-water smaller bays and estuaries of three major rivers with
a wide shelf and steep continental slope. The water exchange between the bay and the
open sea is governed mainly by a cyclonic circulation over the deep central basin
and the Primorskoye current flowing to the southwest along the continental slope
of the Primorsky Krai (Russia).
We have used velocity data from the MHI ocean circulation model \cite{FAO}
which is a set of 3D primitive equations in $Z$-coordinate system with
10 quasi-isopycnal layers and the resolution of 2.5~km.

\begin{figure}[!htb]
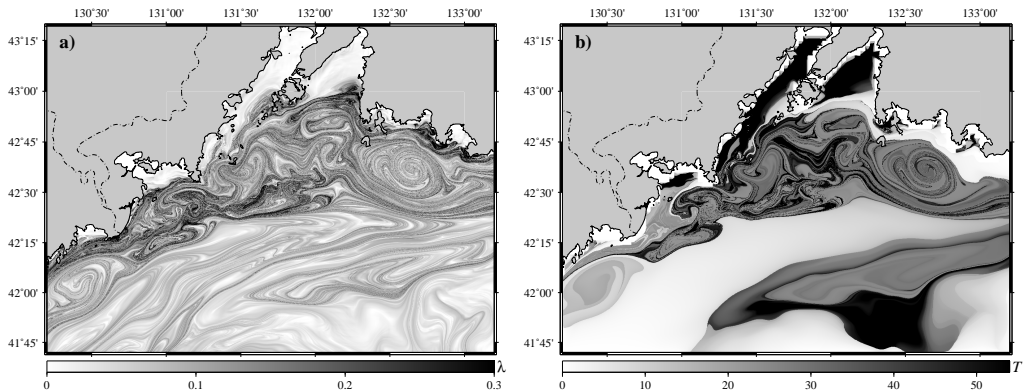

\begin{center}
\includegraphics[width=0.49\textwidth,clip]{fig1a.eps}
\includegraphics[width=0.49\textwidth,clip]{fig1b.eps}
\end{center}
\caption{(a) The Lyapunov map in the Peter the Great Bay
and the surrounding region of the Japan Sea.
(b) The exit-time map in the same region.}
\label{fig1}
\end{figure}

To characterize the water exchange between the Peter the Great Bay and the
open sea we compute the FTLE  map and the exit-time
map (Fig.~\ref{fig1}). A large number of synthetic particles have been uniformly
distributed over the region with [$130^{\circ}12^{\prime}:133^{\circ}12^{\prime}$]~E
and [$41^{\circ}42^{\prime}:43^{\circ}19^{\prime}$]~N. In Fig.~\ref{fig1}a we compute
the FTLE,  $\lambda$, by the method proposed in Ref.~\cite{OM}.
The advection equations (\ref{adveq}) have been integrated forward in time
for 54 days in the August and September of a typical year. The gray shades code
the magnitude of $\lambda$. The value $\lambda=0.085$, at which the distance
between neighbouring particles increases in 100 times, is chosed to be a threshold.
The regions with $\lambda<0.085$ are supposed to be regular, the ones
with $\lambda>0.085$ --- chaotic. The black ridges with $\lambda \gg 0.085$
visualize stable manifolds of influencial hyperbolic trajectories in the region.
Spiral-like structures reveal eddies of different scales, the white and light-grey
zones are the stagnation regions or shear currents. The sandwich-like
structures are signs of the most intense mixing. The synoptic Lyapunov map in
Fig.~\ref{fig1}a shows the scalar filed of this quantity in geographic
coordinates which are initial positions of the synthetic particles. This map along
with the Lyapunov map, computed backward in time (not shown), demonstrates
with a high resolution the complicated character of transport and mixing in
the Peter the Great Bay.

The exit-time map is shown in Fig.~\ref{fig1}b.
The color in the map codes the time, $T$, particles (initially distributed over the same region)
need to reach the open sea or the coastline. In fact, we compute the trajectories till they
reach the 3~km band along the coastline. The white wide band along the coast in Fig.~\ref{fig1}b
demonstrates the Primorskoye current along which particles quickly leave the bay
to the southwest. The large white corridor in the central part of the region selected
separates the Peter the Great Bay from the open sea. Black color marks the particles
that did not leave the bay for the computation time, 54 days. The stagnation zones are situted,
as expected, in the smaller bays, the Amursky and the Ussyrisky ones, which
are visible as black spots on the both sides of the peninsula in the north.
The exit-time map reveals the complicated process
of chaotic mixing in the central part of the bay with the spiral-like
anticyclonic eddy (with the center at $132^{\circ}45^{\prime}$~E and
$42^{\circ}40^{\prime}$~E) and gives a valuable information about origin and fate
of waters.

\begin{figure}[!htb]
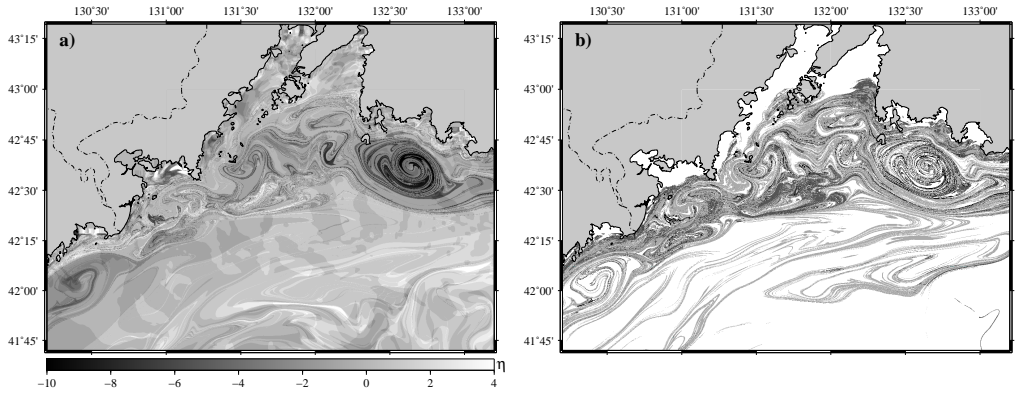

\begin{center}
\includegraphics[width=0.49\textwidth,clip]{fig2a.eps}
\includegraphics[width=0.49\textwidth,clip]{fig2b.eps}
\end{center}
\caption{(a) Rotation and  (b) mixing maps in the Peter the Great Bay.}
\label{fig2}
\end{figure}

To get an information about the character of motion of different waters,
their drift, rotation and oscillation, we compute the new Lagrangian
synoptic maps: rotation and mixing maps, transport and visitor maps.
We compute for a large number of particles the number of cyclonic, $\eta_c$,
and anticyclonic, $\eta_a$, rotations and their difference $\eta$.
The typical kinds of particle's motion are the following: 1) simple drift or
linear displacement if
$\eta_c$, $\eta_a$, $|\eta|$ $<\eta_{\rm cr}=5$, where $\eta_{\rm cr}$ is a
threshold value of the rotation number; 2) rotation, if $|\eta| >\eta_{\rm cr}=5$;
3) oscillation, if $\eta_c$, $\eta_a >\eta_{\rm cr}=5$ but $|\eta|<\eta_{\rm cr}$.
In the rotation map in Fig.~\ref{fig2}a white and black colors mean
cyclonic, $\eta_c$, and anticyclonic, $\eta_a$, rotations, respectively, computed
for the same period of time, 54 days.  Grey color codes the particle with
predominant displacements or oscillations. The map demonstrates clearly
the same spiral-like anticyclonic eddy as in Fig.~\ref{fig1} and the large-scale
filaments with foldings typical to chaotic advection in the ocean.

To characterize the chaotic mixing more clearly we compute along with
the rotation numbers the FTLE $\lambda$. If
$\lambda> \lambda_{\rm cr}=0.85$ and $\eta_a >\eta_{\rm cr}=5$, we will speak
about unstable rotations in the corresponding region. If
$\lambda> \lambda_{\rm cr}=0.85$ but $\eta_a <\eta_{\rm cr}=5$ one deals with
unstable linear displacement of the corresponding particles. The mixing map
in Fig.~\ref{fig2}b shows by color regions with different dynamical properties
specified by the rotation numbers and the maximal Lyapunov exponent. White color
marks the regions with regular oscillations and/or predominant displacements.
The spots of particles, placed in those regions, move as whole being deformed
slightly. The white grey color --- the regions with unstable displacements which
are peripheries of the anticyclonic eddies and their filaments.
The spots, placed in those regions, are elongated strongly. The dark grey color
--- the regions with unstable oscillator motion with the particles rotating
for 54 days in the cyclonic and then in the anticyclonic directions. The black color
corresponds to the unstable rotation that manifests itself in narrow
filaments and spiral-like structures in  anticyclones.

\begin{figure}[!htb]
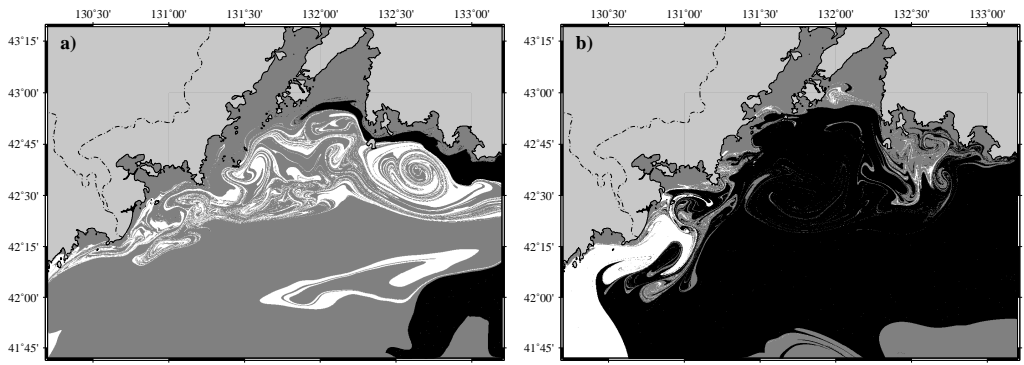

\begin{center}
\includegraphics[width=0.49\textwidth,clip]{fig3a.eps}
\includegraphics[width=0.49\textwidth,clip]{fig3b.eps}
\end{center}
\caption{Transport maps in the Peter the Great Bay.
(a) The particle's fate and (b) origin maps.}
\label{fig3}
\end{figure}

In order to find frontal zones and transport pathways we propose to compute
the transport maps showing the final positions of particles when integrating
the advection equations (\ref{adveq}) forward and backward in time
(see Figs.~\ref{fig3}a and \ref{fig3}b, respectively). In other words,
the equations (\ref{adveq}) have been solved for each of the million particles
initially distributed over the region selected for 54 days
forward and backward in time. In the first case we get the particle's fate map
(Fig.~\ref{fig3}a) with the black (white) particles leaving the bay through
the eastern (western) border. The grey particles are those that did not
leave the bay for the computation time. When integrating
the equations (\ref{adveq}) backward in time, we get the particle's origin map
with the black (white) particles entering the bay through
the eastern (western) border and the resident particles shown in grey.
The frontal zone, separating the waters with different fate and origin,
consistes of smooth, meandered and spiral-like fragments.

\section{Transport and mixing in the Kuroshio Extension region}

The Kuroshio Extension prolongs the Kuroshio Current when
the latter separates from the continental shelf
at about $30^{\circ}$~N. It flows eastward from this point as
a strong unstable meandering jet constituting a front separating the
warm subtropical and cold subpolar waters of the North
Pacific Ocean. It is a region with one of the most intense
air--sea heat exchange and the highest eddy kinetic energy level
strongly affecting climate. Transport of water masses
is of cruicial importance and may cause heating and freshing
of waters with a great impact on the weather and living organisms.

The surface ocean currents used in this section are derived from
satellite altimeter measurements of sea height (http://www.aviso.oceanobs.com).
The velocity data covers
the period from 1992 to 2011 with weekly data on a $1/3^{\circ}$ Mercator grid.
In our study we focus on the region between $30^{\circ}$ and $45^{\circ}$~N and
between $130^{\circ}$ and $165^{\circ}$~E. Bicubical spatial interpolation
and third order Lagrangian polinomials in time have been used to provide
accurate numerical results. Lagrangian synoptic maps, manifolds and chaotic advection
structures in general are determined by the large-scale advection field,
which is appropriately captured by altimetry.
Thus, computation of particle's trajectories statistically is not especially sensitive to
imperfections of the velocity field caused by the interpolation and measurement
imperfections.

\begin{figure}[!htb]
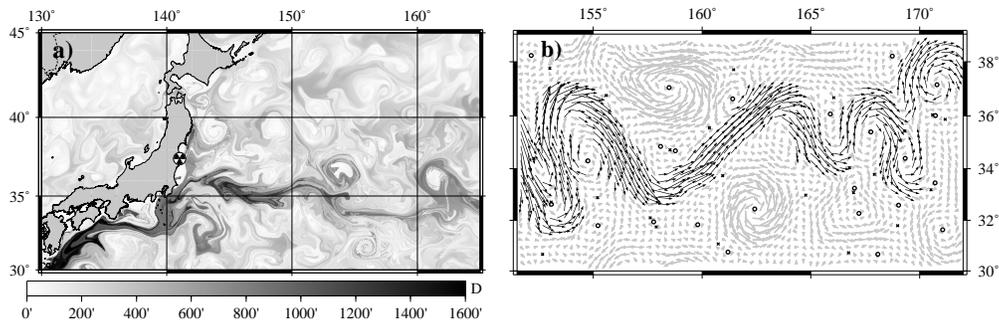

\begin{center}
\includegraphics[width=0.47\textwidth,clip]{fig4a.eps}
\includegraphics[width=0.49\textwidth,clip]{fig4b.eps}
\end{center}
\caption{(a) Displacement map in the Kuroshio Extension region.
The color codes the magnitude of displacement $D$ in
minutes. The Fukushima Daiichi nuclear power plant
is marked with the sign of radioactivity. (b) Velocity field of the region on
1 January, 2010. Circles and crosses are instantaneous elliptic and hyperbolic
points, respectively.}
\label{fig4}
\end{figure}

In Fig.~\ref{fig4}a we demonstrate the displacement map for the region computed
for 45 days after the beginning of the incident at the Fukushima Daiichi
nuclear power plant.
The shades of gray depict the magnitude of the
displacement of a tracer $D=\sqrt{(x_f-x_0)^2 + (y_f-y_0)^2}$,
from its initial position, $(x_0, y_0)$, to a final one $(x_f, y_f)$.
The Kuroshio Current is well pronounced including meanders and
intrusions, its extension, and mesoscale eddies. Two
light-colored eddy patches are of particular interest.
Their centers are approximately at the latitude of the
Fukushima plant and at longitudes $153^{\circ}$~E
(the mushroom-like dipolar eddy) and $161^{\circ}$~E
(the circular eddy), both eddies being surrounded by dark-colored necklaces
having a relatively high magnitude of $D$. This pattern
exemplifies the ring birth process due to the
meandering of the Kuroshio current and subsequent
detachment of eddies from the main jet.

To get a picture of an ``instantaneous '' state of the region we show
in Fig.~\ref{fig4}b the surface velocity field computed on the fixed day, 1 January,
2010. The main meandering jet is depicted by the black arrows corresponding
to comparatively large velocities. The grey arrows around the jet with smaller
velocities reveal a number of cyclonic and anticyclonic eddies on both
flanks of the jet. We have computed instantaneous elliptic and hyperbolic
points of the flow and showed them by circles and crosses, respectively.
The elliptic points are situated mainly in the centers of the eddies, whereas
the hyperbolic ones are in the regions between the eddies with different polarity
and in the periphery of isolated eddies. The hyperbolic points are especially
important because they may be connected by instantaneous stable and unstable
manifolds dividing the flow into regions with cardinally different dynamics.

\begin{figure}[!htb]
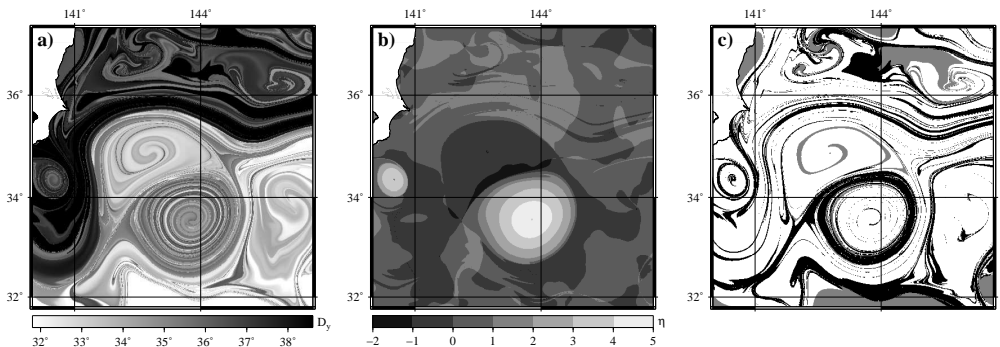

\begin{center}
\includegraphics[width=0.32\textwidth,clip]{fig5a.eps}
\includegraphics[width=0.32\textwidth,clip]{fig5b.eps}
\includegraphics[width=0.32\textwidth,clip]{fig5c.eps}
\end{center}
\caption{Lagrangian maps of the AC--C vortex pair in which the color codes:
(a) the meridional displacement of synthetic particles, $D_y$,
on the 60th day of integration, (b) the number of their rotation around the
vortex centers
on the 15th day and (c) their Lyapunov exponents, $\lambda$, and, $D$,
displacements on the 45th day with
the following legenda:
white means $\lambda <\lambda_{\rm cr}$, $D \ge D_{\rm cr}$,
grey --- $\lambda <\lambda_{\rm cr}$, $D<D_{\rm cr}$
and black --- $\lambda \ge \lambda_{\rm cr}$, $D \ge D_{\rm cr}$.}
\label{fig5}
\end{figure}

We plan to show in this section that the Lagrangian diagnostics is well
suitable to describe the mesoscale and submesoclace features of
the complex picture of mixing in the Kuroshio Extension region.
The altimetric velocity data we used covers
the period from 1 January to 3 July, 2010.  We focus on a vortex pair on
the jet's southern flank, consisting of the anticyclone (AC) and cyclone
(C). The pair manifests itself on the Lagrangian maps in
Fig.~\ref{fig5} computed for a large number of synthetic particles seeded over the region
considered. All the maps visualize the northern hat-like AC
with the axes of 150 and 100~km and the southern circular C with the diameter
150~km. In Fig.~\ref{fig5}a the color codes the meridional displacement, $D_y$,
of particles on the 60th day of integration. The spiral structure of
the C is well developed with the spiral untwisting counter-clockwise, whereas
it is less pronounced for the AC with the spiral untwisting clockwise.
The character of the water motion in the C and AC is also different and
becomes evident after computing
the number of particle's rotation around the vortex centers. It follows from
Fig.~\ref{fig5}b that water in the AC core circulates with approximately the same
angular velocity, whereas this quantity decreases from the center of the C to
its periphery (pay attention to the ring-like structure of the C). In order
to visualize the stable manifolds of the hyperbolic trajectories around
the vortex pair, we compute in Fig.~\ref{fig5}c the FTLE, $\lambda$, and
displacements, $D$, of the particles. The shades of grey in this figure
modulate different combinations and magnitudes of $\lambda$ and $D$ with
respect to some chosen ``critical'' values: $\lambda_{\rm cr}$, corresponding to
divergence of initially close particles over 100~km, and $D_{\rm cr}=100$~km.
The black convoluted curves in the figure between the eddies, around each
of them and around the very pair delineate the corresponding $W_s$ manifolds.

\begin{figure}[!htb]
\begin{center}
\includegraphics[width=0.98\textwidth,clip]{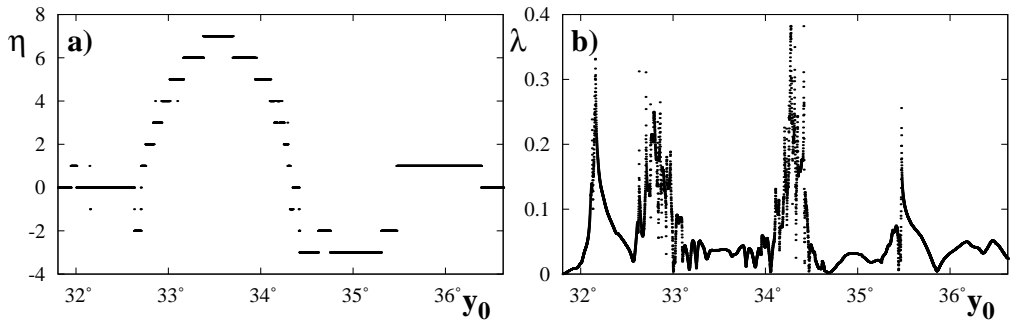}
\end{center}
\caption{The scattering plots for the vortex pair on the 30th day of integration.
(a) Number of times, $\eta$, the particles rotate around the vortex centers
vs initial particle's latitude position $y_0$, (b) the corresponding maximal
FTLE vs $y_0$.}
\label{fig6}
\end{figure}

To give a detailed description of the structure of each eddy in the vortex
pair we apply the method of particle's scattering elaborated in Ref.~\cite{BUP04}.
We cross both the eddies by a material line and compute
rotation number $\eta$, and the maximal FTLE on initial particle's
latitude $y_0$. The scattering plot in Fig.~\ref{fig6}a demonstrates that the
waters in the C core really rotate with different angular velocities decreasing
from the center to its periphery. Rotation in the AC core is much more homogeneous.
Moreover, the waters in the C rotate in two times
faster than in the AC. The scattering  plot $\lambda (y_0)$ in Fig.~\ref{fig6}b
demonstrates smooth segments in the cores of C and AC and irregular oscillations
in their periphery. It simply means that the water in the cores moves more or
less coherently whereas the motion in the eddy's peripheries is erratic due
to numerous intersections of stable and unstable manifolds. Computation of
the dependence of the time of exit of the particles $T$,
belonging to the material line, on $y_0$ (not shown) confirms that
waters prefer to quit the C more or less periodically by portions.
Each portion is represented by a $\cup$-like segment of the $T(y_0)$ function
which consists of a large
number of particles with approximately the same time of exit and the
same rotation number $\eta$. In difference from the C, particles quit the AC
core practically at the same time. In other words, the particles quit the C
by portions along spiral-like transport pathways, whereas the periphery of the
AC exchanges water with the surrounding but its core
moves coherently as a whole for a time.

\begin{figure}[!htb]
\begin{center}
\includegraphics[width=0.68\textwidth,clip]{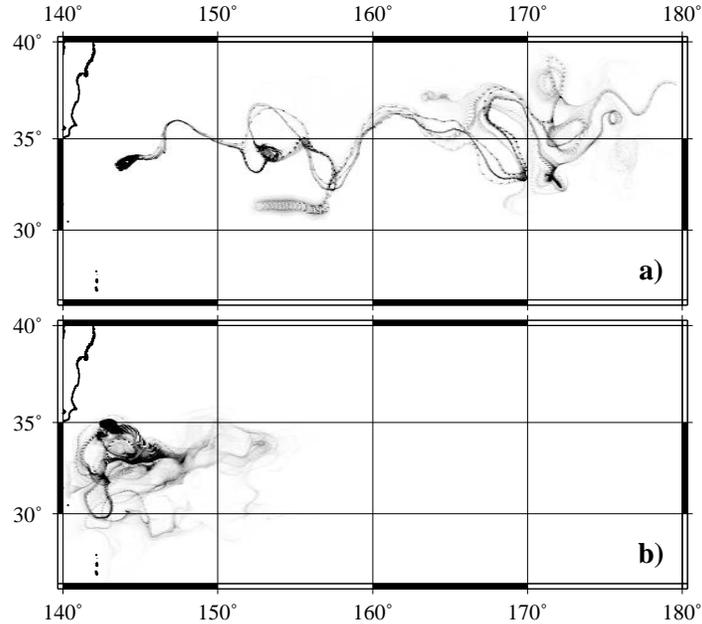}
\end{center}
\caption{Visitor maps for (a) the cyclone and (b) anticyclone show how
frequently fluid particles from the corresponding eddy's cores visit
for 180 days different places in the Kuroshio Extension region.}
\label{fig7}
\end{figure}

In conclusion we demonstrate in Fig.~\ref{fig7} how
frequently fluid particles, chosen in the cores of the C and AC,
visit for 180 days different places in the Kuroshio Extension region.
It is evident that the C was absorbed by the main jet in a short time  and
then its waters travelled eratically within the jet with a few excursions to
its northern and southern flanks. It is interesting that in course of time
C waters have formed the new cyclonic eddy nearby
$(x_0=155^{\circ}$~E, $y_ 0=32^{\circ}$~N).
In contrast to the C, the AC waters have walked eratically on the southern
flank of the jet in a restricted region within $x_0=[140^{\circ}: 150^{\circ}]$~E,
$y_ 0=[28^{\circ}:35^{\circ}]$~N.

\section{Conclusion}

The Lagrangian approach has been shown to be very useful to gain new information
on chaotic transport and mixing in the ocean. We have elaborated
new Lagrangian diagnostic tools to visualize and quantify those processes: the time of exit of
fluid particles off a selected
box, their displacements, the number of their cyclonic and anticyclonic
rotations and the number of times they visit different places in the region.
Along with the Lyapunov maps, the corresponding high-resolution Lagrangian
synoptic maps of
those quantities, computed by solving advection equations forward and
backward in time for different periods of the year, are new diagnostic and
prognostic products characterizing the state of the ocean.
The technique developed can be applied to the global ocean and its basins.

In this paper we have focused on a comparatively small
marine bay, the Peter the Great Bay in the Japan Sea near Vladivostok (Russia),
and on a comparatively large region in the North Pacific, the Kuroshio Extension
system. In the bay study in summer and autumn periods, we have used
the velocity data from a Japan Sea
eddy-resolved circulation numerical model with the resolution of 2.5~km.
It has been shown that the Lyapunov and exit-time maps, the rotation,
mixing and transport maps allowed to quantify and specify movement of
water masses, their mixing and the degree of its chaoticity in the bay.
Those high-resolution maps allowed to visualize transport pathways by
which waters exit and enter the bay.

As to the Kuroshio Extension, we have used the velocity data
derived from satellite altimeter measurements of sea height with the corresponding
interpolation. The main attention has been paid to study structure,
transport and mixing of a vortex pair with strongly interacting cyclonic and anticyclonic
eddies. Such dipoles occur frequently in that region.
We have computed Lagrangian synoptic maps
for the time of exit of particles,
the number of changes of the sign of zonal and meridional velocities,
and for other quantities. Along with the Lyapunov
map, they have been shown to be able to reveal the vortex structure and its
evolution,
meso- and  submesoscale filaments, repelling material lines, hyperbolic and
non-hyperbolic regions in the sea. In particular, we have found that the
eddies have a prominent spiral-like structure resembling the spiral
patterns at satellite images in that region.

The work was supported by the Program
``Fundamental Problems of  Nonlinear Dynamics'' of the Russian
Academy of Sciences, by the Russian Foundation
for Basic Research (projects nos. 09-05-98520 and 11-01-12057)
and by the
Prezidium of the Far-Eastern Branch of the RAS.

\bibliographystyle{amsplain}
\bibliography{ws-pro-sample}

\providecommand{\bysame}{\leavevmode\hbox to3em{\hrulefill}\thinspace}
\providecommand{\MR}{\relax\ifhmode\unskip\space\fi MR }
\providecommand{\MRhref}[2]{%
  \href{http://www.ams.org/mathscinet-getitem?mr=#1}{#2}
}
\providecommand{\href}[2]{#2}
\begin{thebibliography}{10}

\bibitem{BUP04}
M.~V. Budyansky, M.~Yu. Uleysky, and S.~V. Prants, JETP \textbf{99} (2004),
  no.~5, 1018--1027.

\bibitem{OI09}
Francesco d'Ovidio, Jordi Isern-Fontanet, Crist{\'o}bal L{\'o}pez, Emilio
  Hern{\'a}ndez-Garc{\'i}a, and Emilio Garc{\'i}a-Ladona, Deep Sea Research
  Part I: Oceanographic Research Papers \textbf{56} (2009), no.~1, 15--31.

\bibitem{H02}
G.~Haller, Physics of Fluids \textbf{14} (2002), no.~6, 1851--1861.

\bibitem{Haller}
G.~Haller and A.C. Poje, Physica D: Nonlinear Phenomena \textbf{119} (1998),
  no.~3-4, 352--380.

\bibitem{KP06}
K.V. Koshel and S.V. Prants, Physics Uspekhi \textbf{49} (2006), 1151--1178.

\bibitem{LC05}
Francois Lekien, Chad Coulliette, Arthur~J. Mariano, Edward~H. Ryan, Lynn~K.
  Shay, George Haller, and Jerry Marsden, Physica D: Nonlinear Phenomena
  \textbf{210} (2005), no.~1-2, 1--20.

\bibitem{MS06}
Ana~M. Mancho, Des Small, and Stephen Wiggins, Physics Reports \textbf{437}
  (2006), no.~3-4, 55--124.

\bibitem{Ottino}
J.M. Ottino, \emph{The kinematics of mixing: Stretching, chaos, and transport},
  Cambridge University Press, Cambridge, U.K., 1989.

\bibitem{OM}
S.~V. Prants, M.~V. Budyansky, V.~I. Ponomarev, and M.~Yu. Uleysky, Ocean
  modelling \textbf{38} (2011), no.~1-2, 114--125.

\bibitem{FAO}
S.~V. Prants, V.~I. Ponomarev, M.~V. Budyansky, M.~Yu. Uleysky, and P.~A.
  Fayman, Izvestiya, Atmospheric and Oceanic Physics (in press).

\bibitem{PNAS}
Emilie Tew~Kai, Vincent Rossi, Joel Sudre, Henri Weimerskirch, Cristobal Lopez,
  Emilio Hernandez-Garcia, Francis Marsac, and Veronique Gar\c{c}on, PNAS
  \textbf{106} (2009), no.~20, 8245--8250.

\bibitem{WA06}
Darryn~W. Waugh, Edward~R. Abraham, and Melissa~M. Bowen, Journal of Physical
  Oceanography \textbf{36} (2006), no.~3, 526--542.

\end{thebibliography}

\end{document}